\documentclass[pra,twocolumn,showpacs]{revtex4}

\usepackage{epsfig}
\usepackage{graphicx}
\usepackage{bm}
\usepackage{amsmath}
\usepackage{dcolumn}

\newcommand{\rcm}{\mbox{cm$^{-1}$}}
\newcommand{\Sch}{Schr\"{o}dinger}

\begin{document}

\title{The potential of the ground state of NaRb}

\author{O. Docenko}
\author{M. Tamanis}
\author{R. Ferber}
\affiliation{Department of Physics, University of Latvia, 19 Rainis boulevard, Riga LV-1586, Latvia}
\author{A. Pashov}\altaffiliation{On leave from the Institute for Scientific Research in Telecommunications,
 ul. Hajdushka poliana 8, 1612 Sofia, Bulgaria}
\author{H. Kn\"ockel}
\author{E. Tiemann}
\affiliation{Institut f\"ur Quantenoptk, Universit\"at Hannover,
  Welfengarten 1, 30167 Hannover, Germany}

\date{\today}

\begin{abstract}
The X$^{1}\Sigma ^{+}$ state of NaRb was studied by Fourier transform spectroscopy.
An accurate potential energy curve was derived from more than
8800 transitions in isotopomers $^{23}$Na$^{85}$Rb and $^{23}$Na$^{87}$Rb.
This potential reproduces
the experimental observations within their uncertainties of 0.003 \rcm\
to 0.007 \rcm. The outer classical turning point of the last observed energy
level ($v''=76$, $J''=27$) lies at $\approx 12.4$ \AA, leading to a energy of
4.5 \rcm\ below the ground state asymptote.
\end{abstract}


  \maketitle

\section{Introduction}
 \label{intro}

The heteronuclear alkali dimers attract interest of both
experimental and theoretical researchers involved in collision
dynamics, photoassociative spectroscopy, laser cooling and
trapping of alkali atoms \cite{Shaffer:99,Ferrari:02,Hadzibabic:02,Weiss:03}.
Special interest is put on the study of the ground states and especially near
the atomic asymptote, since the precise knowledge of the long-range
interactions between two different types of alkali atoms
is necessary for understanding and realization of such cold
collision processes as sympathetic cooling,
formation of two species BEC (TBEC) and ultracold heteronuclear molecules.

Although the NaRb molecule is one of the promising candidates for
TBEC \cite{Weiss:03}, the experimental information on the ground singlet
state is still limited.

Accurate term values of the X$^{1}\Sigma ^{+}$ state were determined
by the Doppler-free polarization spectroscopy \cite {Wang2:91}.
They cover a range of low vibrational quantum numbers $v'' \in \lbrack 0,6]$
with limited sets of rotational quantum numbers (e.g. $J'' \in [1,80]$ for
$v''=0$ and $J'' \in [18,36]$ for $v''=6$). Later  Kasahara et al.
\cite{Kasahara:96} extended the range of $v''$ up to $v''=30$, but only for
$J''=10,12$.
On the other hand the recent experiment
by the Riga group \cite{Docenko:02} collected about 300 transition
frequencies to energy levels with a much wider spectrum of vibrational and rotational
quantum numbers ($v''\in [15,76]$, $J''\in [12,64]$) but with
relatively moderate accuracy (~0.1 \rcm~). A hybrid potential
for the NaRb ground state  based
on extrapolation between regions determined from the early
experimental data \cite{Wang2:91,Kasahara:96}
and the theoretical dispersion coefficients from
Ref. \cite{Marinescu:99} was constructed
by Zemke et al. \cite{Zemke:NaRb} giving an improved estimate on the dissociation
energy of the ground state - $D_{\mathrm e}=5030.75$ \rcm.

\begin{figure}
  \centering
 \epsfig{file=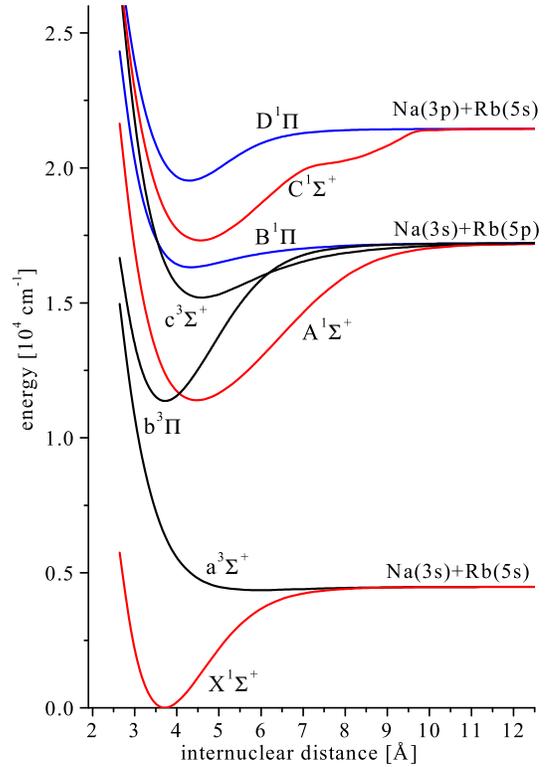,width=0.9\linewidth}
  \caption{Selected low singlet and triplet electronic states in NaRb
according to Ref.~\cite{Korek:00}.}
 \label{potfr}
\end{figure}

On the theoretical side two papers have recently appeared: by Korek et al.
\cite{Korek:00} and by Zaitsevskii et al. \cite{Zaitsevskii:01}.
In both of them theoretical potential energy curves for the ground
and several excited states are given. Results from Ref.~\cite{Korek:00}
on selected low electronic states are shown also in Figure~\ref{potfr}, which will help in
following the electronic assignment of the new observations, reported below.

The goal of the present investigation is to study extensively
the ground electronic state X$^{1}\Sigma ^{+}$  of the NaRb molecule.
We chose the Fourier transform spectroscopy which is able to provide
abundant and precise spectroscopic data, needed for construction of
an accurate potential energy curve. In addition, this study is
considered as the necessary background for further investigations of this
molecule at long internuclear distances required for modeling cold collisions
at 1mK and below.

\section{Experiment}
 \label{exper}

The NaRb molecules were produced in a single section heat-pipe oven
similar to that described in Ref.~\cite{Allard:02} by heating 5 g of Na
(purity 99.95 \%) and 10 g of Rb
(purity 99.75 \%, natural isotopic composition) from Alfa Aesar.
The oven was operated at temperatures between 560 K and 600 K
and typically with 2 mbar of Ar as buffer gas.
At these conditions apart from the atomic vapors all three
types of molecules are formed - Na$_2$, Rb$_2$ and NaRb.
This mixture was illuminated by three different laser sources:
a single mode, frequency doubled Nd:YAG laser, Ar$^+$ ion laser
and Rhodamine 6G dye laser. Their wavelengths fall in a spectral
window with weak absorption of Rb$_2$. Setting the working temperatures
as low as possible we reduced the density of the Na$_2$ molecules reaching
a relatively intense NaRb spectra free of Na$_2$ emission.
The heat pipe oven was in operation more than 60 hours without
refilling and at the end of the experiment was still in good working
condition.

The fluorescence from the oven was collected in the direction
opposite to the propagation of the laser beam and then focused
into the input aperture of Bruker IFS 120 HR Fourier transform
spectrometer. The signal was detected with a Hamamatsu
R928 photomultiplier tube or a silicon diode. The scanning
path of the spectrometer was set to reach a typical resolution of
0.0115 \rcm\ $\div$ 0.03 \rcm\ and
the number of scans for averaging varied between 10 and 20. In order to avoid the
illumination of the detector by the He-Ne laser, used for calibration of the spectrometer,
a notch filter with 8 nm FWHM was introduced. For better signal-to-noise
ratio some spectra were recorded by limiting the desired spectral
window with colored glass or interference filters.

The single mode, frequency doubled Nd:YAG laser with a typical output
power of 70 mW at 532.1 nm excited the C$^1\Sigma^+ \leftarrow$ X$^1\Sigma^+$
and D$^1\Pi \leftarrow$ X$^1\Sigma^+$
systems of NaRb, compare Fig. 1. The frequency was varied between 18787.25 \rcm\ and
18788.44 \rcm\ and spectra were recorded at frequencies which excite strong
fluorescence.

\begin{figure}
  \centering
 \epsfig{file=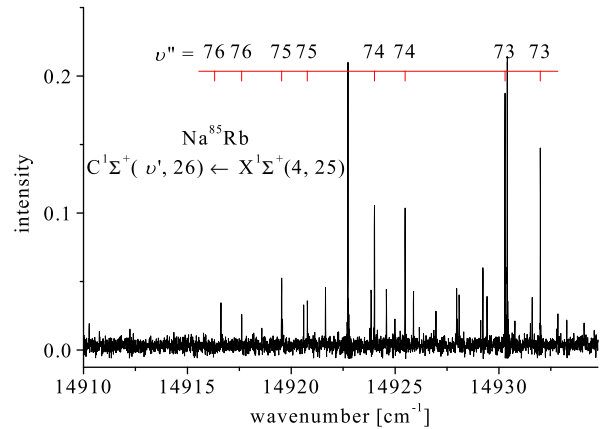,width=\linewidth}
  \caption{The vibrational progression up to $v''=76$ in Na$^{85}$Rb
   excited by a single mode Ar$^+$ laser. Lines not indicated
 are also assigned.}
 \label{v76}
\end{figure}

The Ar$^+$ ion laser was operated both in single (typical power 100$\div$500 mW)
and multi mode (typical power 0.5$\div$3 W) regimes.
The 514.5 nm, 501.7 nm, 496.5 nm, 488.0 nm and 476.5 nm Ar$^+$ ion laser lines  
induced fluorescence mainly due to the D$^1\Pi \leftarrow$ X$^1\Sigma^+$
system of NaRb. The Na$_2$ B$^1\Pi_{\mathrm u} \rightarrow$ X$^1\Sigma^+_{\mathrm g}$
band was also observed, especially exciting with the bluer Ar$^+$ laser lines, but
its intensity was reduced by decreasing the working temperature down to
560 K. Along with the D $\leftarrow$ X system, the 514.5 nm line excites also
transitions in the  C$^1\Sigma^+\leftarrow$ X$^1\Sigma^+$
system of NaRb  \cite{Docenko:02}. The special form of potential energy curve
of the C$^1\Sigma^+$ state \cite{Korek:00} results in favorable Franck-Condon factors for
transitions from some C state levels both to low and high lying energy levels of the
ground state. Such C state levels are excited by the 514.5 nm line and
the green florescence of the D $\leftarrow$ X system is accompanied by
a red fluorescence due to C $\leftarrow$ X transitions, which reach $v''=76$
of the ground state (Fig.~\ref{v76}).

In order to bring the frequency of the single mode Ar$^+$
laser exactly in resonance with desired transitions we
did the following. As a first step we recorded the LIF for a given
frequency simultaneously with the Fourier spectrometer and a
GCA/McPherson Instruments scanning monochromator (1m focal length). The high resolution
Fourier spectrum helped us to assign unambiguously the record of
the monochromator. Then we set the monochromator on a strong and
pronounced line of the progression of interest and we tuned the
Ar$^+$ laser frequency until the maximum signal on
the monochromator was reached. The frequency stability of the laser 
with a temperature stabilized intracavity etalon was sufficient 
in order to perform 10 to 20 scans with the Fourier spectrometer.

\begin{figure}
  \centering
 \epsfig{file=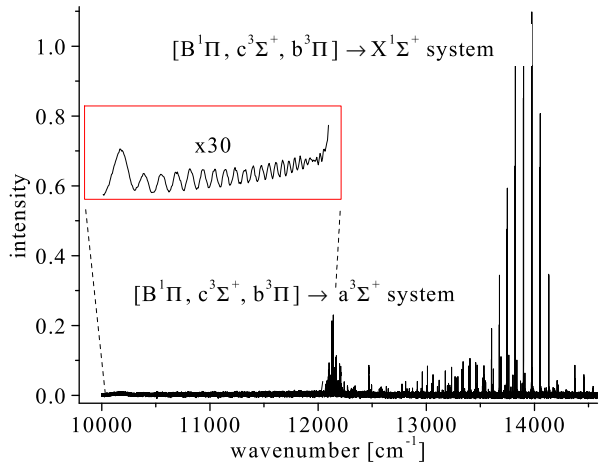,width=\linewidth}
  \caption{The [B$^1\Pi$, b$^3\Pi$, c$^3\Sigma^+$]-X$^1\Sigma^+$ and
 [B$^1\Pi$, b$^3\Pi$, c$^3\Sigma^+$]-a$^3\Sigma^+$
  bands of NaRb excited by a single mode Rhodamine 6G laser. The insert shows the
  continuum due to the bound-free emission to the repulsive wall of the
 a$^3\Sigma^+$ state.}
 \label{b-x}
\end{figure}

The Rhodamine 6G laser excites the B$^1\Pi \leftarrow$ X$^1\Sigma^+$
transition and also weak transitions in the C$^1\Sigma^+ \leftarrow$ X$^1\Sigma^+$
system and was used for two reasons. The first was to enrich the
information on the ground state levels with intermediate vibrational
quantum numbers because of the gap between the levels from the
D$^1\Pi \leftarrow$ X$^1\Sigma^+$ system (mainly with low $v''$) and those
from the C$^1\Sigma^+ \leftarrow$ X$^1\Sigma^+$  system
(mainly with high $v''$). Secondly, we wanted to excite levels of the B state
with significant triplet admixture due to perturbations from the
neighboring b$^3\Pi$ and c$^3\Sigma^+$ states \cite{Wang2:91,Wang:92}.
Indeed, scanning the frequency of the dye laser from 16729 \rcm\ to 16965 \rcm\
we encountered
a large number of excitations where a second fluorescence  band around 12000 \rcm\
appeared along with the B $\leftarrow$ X system (see Fig.~\ref{b-x}).
The analysis of this band
(a report is in preparation) confirmed that we
have observed transitions to the ground triplet state a$^3\Sigma^+$.

All transitions excited in Na$^{85}$Rb and Na$^{87}$Rb by
the single mode Nd:YAG and Rhodamine 6G lasers
and  by the Ar$^+$ laser in single- and multi mode, assigned and used in this analysis
are listed in Table~1 of the supplementary material (EPAPS No. ....).

 \section{Analysis}
 \label{analysis}

\begin{figure}
  \centering
 \epsfig{file=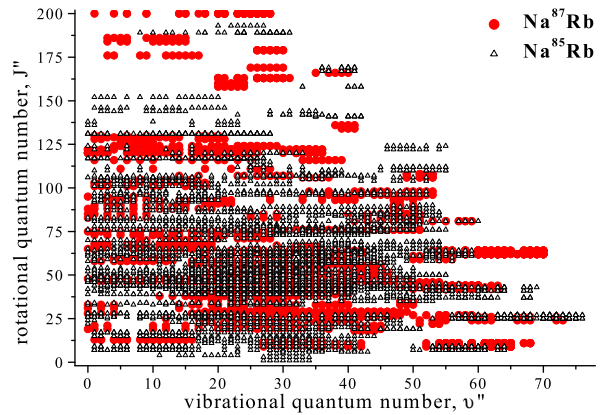,width=\linewidth}
  \caption{The range of vibrational and rotational quantum numbers
of the observed ground state energy levels in Na$^{85}$Rb and Na$^{87}$Rb}\label{dataset}
\end{figure}

The assignment of the recorded spectra was simplified by the
published potential energy curve in Ref.~\cite{Docenko:02}.
After we identified the strongest progressions, which were already
observed in previous studies \cite{Takahashi:81,Nikolayeva:00,Zaitsevskii:01,Docenko:02}, a new potential
was fitted. This preliminary potential was further
improved filling additionally collected experimental data; this sequential procedure allows
continuous checking of the assignment, especially for high $J''$ and in cases
of large gaps in $v''$.
The total data set (Fig.~\ref{dataset})
consists of more than 6150 transitions in Na$^{85}$Rb and 2650
in Na$^{87}$Rb given in Table~2 of the supplementary material (EPAPS No.....).

For fitting the experimental data we used the approach adopted in
Ref. {\cite{Allard:02}}. In order to exclude any fitting parameters concerning
the excited states, we fitted the ground state potential directly to the observed
differences between ground state levels. In our case of 8800 transitions
we selected about 43300 differences for representing the different cases
of vibrational intervals (for details see also \cite{Allard:02}).
The uncertainties of the differences were estimated taking into account the instrumental accuracy
and resolution, and the signal-to-noise ratio for both transitions forming
the difference. The uncertainty of strong lines was set to 0.002~$\div$~0.003~\rcm\
increasing to 0.007 \rcm\ for weaker ones with SNR $\approx 2$. Differences which
were measured several times were averaged with weighting factors
determined from their uncertainties.

As already discussed in \cite{Allard:02}, for our experimental set up
a fluorescence progression excited by a single mode laser will be
shifted in frequency if the laser is not tuned to the center of the Doppler
profile. This problem is partially overcome by fitting differences instead
of transition frequencies. Nevertheless there can be still a residual shift
of the differences due to the frequency dependence of the Doppler shift.
For example if the laser frequency around 20000 \rcm\ is detuned
from the line center by one Doppler width (about 0.03 \rcm\ for
NaRb at 600 K) the difference between two transitions $\nu_1$ and $\nu_2$ separated
by $\nu_1-\nu_2$= 4000 \rcm\ will be shifted from the ``true'' one by 0.006 \rcm.
This shift is comparable with the typically expected uncertainty
($ \approx 0.004$ \rcm) but it cannot be determined in our experiment.
Therefore, initially we did not correct the uncertainties of the large
differences (a possible solution could be to increase the uncertainty of the difference
$\nu_1-\nu_2$ proportionally to the ratio $(\nu_1-\nu_2)/\nu_1$). Large
differences however were checked carefully during the fitting procedure and finally no
need for such correction was found.

\begin{figure}
  \centering
 \epsfig{file=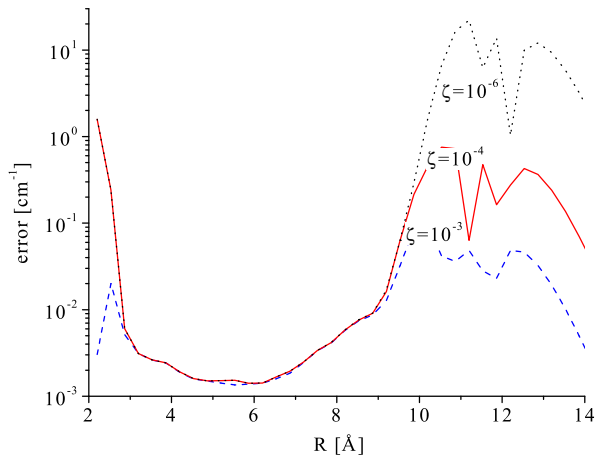,width=\linewidth}
  \caption{The derived standard deviation (in logarithmic scale) of the points of the pointwise
   potential curve for three different values of the singularity parameter $\zeta$
   (for details see Ref.~\cite{Allard:02}).}
 \label{uncert}
\end{figure}

\section{Construction of PEC}
\label{PEC}

The ground singlet states of both isotopomers of NaRb are described
in the adiabatic approximation with a single potential energy curve.
The potential was constructed as a set of points $\{R_i,U(R_i)\}$ 
(see Ref.~\cite{ipaasen}) connected with cubic spline function.
The pointwise representation of the potential,
however, is not convenient for long internuclear distances.
On the other hand, in order to ensure proper boundary 
conditions for solving the \Sch\ equation for high vibrational
energy levels, distances up to $R \approx 20$ \AA\ are needed.
Therefore for $R \ge R_{\mathrm {out}}$ we adopted the usual 
dispersion form:

\begin{equation}
U(R)=D_e-C_6R^{-6}-C_8R^{-8}-C_{10}R^{-10} \mbox{ ,}
\label{lrexp}
\end{equation}

\noindent with coefficients $C_6, C_8$ taken from Ref.~\cite{Derevianko:01a,
Porsev:03}. The connecting point $R_{\mathrm{out}}$ and the 
parameters $D_e$ and $C_{10}$ were varied in order to ensure a 
smooth connection with the pointwise potential as follows.
Initially, the PEC was constructed in a pointwise form up to 16 \AA.
After each fitting iteration $D_e$ and $C_{10}$ were adjusted
to fit the shape of the pointwise potential
between 13.0 \AA\ and 16.0 \AA\ to the analytic formula (\ref{lrexp}).
The crossing point of the pointwise and the long range curves 
was taken as $R_{\mathrm{out}}$. Since
the experimental data are almost insensitive to details of the potential shape 
in this region, we found this procedure sufficient and did not fit the
long range parameters directly to the experimental data. Once the final form of the PEC was achieved,
we found that it is possible to extend the validity of the long range expression
down to approximately 11.8 \AA. At the same time the quality of the derived
potential did not change if the analytic formula (\ref{lrexp})
was fitted to the shape of the pointwise potential between 11.8 \AA\ and
13.24 \AA. A discussion on the advantage of using this matching 
procedure can be found in Ref.~\cite{MC:03}. 

The final set of potential parameters consists of 51 points. In order to calculate the value 
of the potential for $R<R_{\mathrm {out}}$ a natural cubic spline \cite{Numer}
through \emph{all} 51 points and for $R \ge R_{\mathrm {out}}$ the long range expression (\ref{lrexp})
should be used with parameters listed in Table~\ref{ipapot}.
This final PEC gives a standard deviation of $\sigma=0.0031$ \rcm\ and
a normalized standard deviation  of $\bar{\sigma}=0.70$ showing 
the internal consistency of the data set and the quality of the 
derived potential.

\begin{table}

  \centering
  \caption{Pointwise representation of the potential energy curve for
 the X$^1\Sigma^+$ state of NaRb.}
 \begin{tabular*}{0.9\linewidth}{@{\extracolsep{\fill}}rrrr}\hline
 R [\AA] & U [cm$^{-1}$]& R [\AA] & U [cm$^{-1}$]\\ \hline
2.101916 & 21110.5539  &   5.167536 & 2866.7126 \\
2.200000 & 16374.1819  &   5.328985 &    3174.1150 \\
2.298084 & 12751.3674  &   5.490435 &    3450.2078     \\
2.396168 & 9944.6386  &   5.651884 &    3694.1882     \\
2.494252 & 7761.9582  &   5.813333 &    3906.7423     \\
2.592337 & 5992.3235  &   6.065085 &    4179.8842     \\
2.690421 & 4637.9707  &   6.253898 &    4343.3002       \\
2.788505 & 3543.2165  &   6.442712 &    4476.4609       \\
2.886589 & 2638.9307  &   6.631525 &    4584.0755       \\
2.984673 & 1900.4037  &   6.820339 &    4670.5331       \\
3.082757 & 1308.0797  &   7.009152 &    4739.7196       \\
3.180841 & 845.5522   &   7.197966 &    4794.9580       \\
3.278926 & 498.1298   &   7.386780 &    4839.0226       \\
3.377010 & 252.2899   &   7.556326 &    4870.9517       \\
3.475094 & 95.41660   &   7.783673 &    4904.8112        \\
3.573178 & 15.73046   &   8.098462 &    4939.0358         \\
3.671262 & 2.3375     &   8.384103 &    4961.0335        \\
3.792152 & 62.2402    &   8.669744 &    4977.0314        \\
3.933165 & 217.7034   &   9.014483 &    4990.8655        \\
4.074177 & 443.0825   &   9.398621 &    5001.4491        \\
4.215190 & 718.3282   &   9.722105 &    5007.8576        \\
4.360290 & 1035.8028  &  10.201818 &    5014.3657        \\
4.521739 & 1411.4474  &  11.214546 &    5022.0531        \\
4.683188 & 1794.2206  &  12.227273 &    5025.7686      \\
4.844638 & 2170.6796  &  13.240000 &    5027.9612      \\
5.006087 & 2530.6262  &  &                             \\
&&&\\
\multicolumn{2}{l}{$D^{\mathrm X}_{\mathrm e}$=5030.8480 \rcm} & & \\
\multicolumn{2}{l}{$R_{\mathrm{out}}$=12.090520 \AA} &
\multicolumn{2}{l}{$C_8$=3.590$\cdot 10^{8}$ \rcm\AA$^{8}$} \\
\multicolumn{2}{l}{$C_6$=1.293$\cdot 10^{7}$ \rcm\AA$^{6}$} &
   \multicolumn{2}{l}{$C_{10}$=3.539$ \cdot 10^{10}$ \rcm\AA$^{10}$}\\
 \\\hline
\end{tabular*}
\label{ipapot}
\end{table}

In order to find the region where the PEC is unambiguously characterized by
the experimental data we analyzed the uncertainty of the fitting parameters
as described in detail in Ref.~\cite{Li2F:00,Allard:02}. By minimizing
the merit function $\chi^2$ using the Singular Value Decomposition (SVD) method
\cite{Wilkinson:SVD,Numer} it is possible to determine the parameters
on which $\chi^2$ and consequently the quality of the fit
depend only weakly. The fitting parameters in our case are the values
of the potential itself in a preset grid of points.
In Fig.~\ref{uncert} the uncertainties of the fitted potential
for intermediate internuclear distances are presented for three
different values of the singularity parameter $\zeta$ (see \cite{Li2F:00}).
Although the outer classical turning point of the last observed energy
level ($v''=76$, $J''=27$) is 12.4 \AA\, we see that the shape of the
potential energy curve is unambiguously fixed (within the experimental
uncertainty) approximately only between 3.0 \AA\
and  9.8 \AA\ since the uncertainties there almost do not depend on $\zeta$.
Therefore, although the fitted potential energy curve describes all
the experimental data up to $v''=76$, there is no rigorous
way to estimate the uncertainty of its shape beyond 9.8 \AA\
(see also \cite{Allard:02}). The result of Fig.~\ref{uncert} might indicate
that the correlation between the left (R $<$ 3\AA) and right (R $>$ 9.8 \AA)
branch of the potential becomes significant. In this respect the reported
potential in Table~\ref{ipapot} for these regions is only one possibility
which describes our observations within the error limits. From
Table~\ref{ipapot} we derive some parameters which will be useful
when applying the potential for other calculations and spectroscopic
studies. Their values are not rounded in order to keep
consistency with the data from Table~\ref{ipapot}. 

\begin{itemize}
\item equilibrium distance of the potential,\\ $R_{e}=3.643415$ \AA,
\item position of the level with $v''=0$ and $J''=0$,\\ $E_{00}=53.3117$ \rcm\
with respect to potential minimum,
\item the dissociation energy of the ground state, calculated with
respect to $E_{00}$,\\ $D^{\mathrm X}_0$=4977.5363 \rcm.
\end{itemize}

\begin{figure}
  \centering
 \epsfig{file=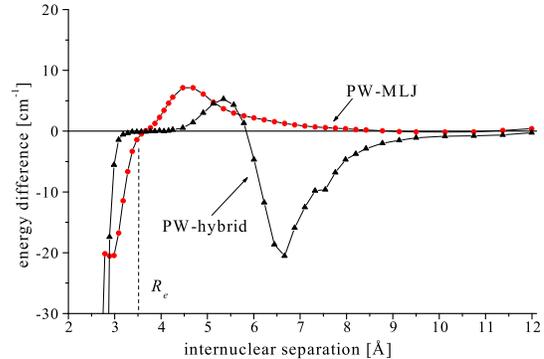,width=0.9\linewidth}
  \caption{Differences between the pointwise representation
of the NaRb ground state potential derived in this study
(PW) and previous potentials energy curves from Ref. \cite{Zemke:NaRb} (Hybrid),
Ref. \cite{Docenko:02} (MLJ). The energy at the equilibrium point $R_e$
is set to zero for all potentials and used as a reference point.}
 \label{compdif}
\end{figure}

In Fig.~\ref{compdif} the potential energy curves published in
Ref.~\cite{Zemke:NaRb,Docenko:02} are compared with the
pointwise potential (PW) of this study. The reason for the
difference between the PW potential and the Modified Lennard-Jones (MLJ)
potential \cite{Docenko:02} around $R_e$ is probably that
for the construction of the later potential the rotational quantum numbers for
levels with $v''<30$ were limited only to the narrow
interval from \cite{Kasahara:96} ($J''=10,12$). With such data set
the rotational constant $B_v$ can be determined only with large uncertainty
which corresponds to shifts of the whole potential along the internuclear
axes which is clearly indicated in Fig.~\ref{compdif}.
As already discussed in Section~\ref{intro} the hybrid potential
from Ref.~\cite{Zemke:NaRb} was constructed using the whole range of
available rotational quantum numbers from the literature
(\cite{Wang2:91, Kasahara:96}) and therefore the agreement with the present
study around $R_e$ is much better. The deviations between the
PW and the hybrid potential reaching 20 \rcm\ for intermediate internuclear distances
are caused by the extrapolation between the theoretical long range part of
the potential and the experimental short range part. In the same region
the experimentally determined shape of the MLJ potential is in a much better
agreement with the present study.

Since the primary data from \cite{Wang2:91} were not available to us,
we were able to check our potential only against the data published in
\cite{Kasahara:96}. Forming differences between transition frequencies
exactly as for our LIF data we found an agreement with the differences
predicted by the present PEC from Table~\ref{ipapot} with a standard deviation
of 0.0026 \rcm, well within the expected experimental error of 100 MHz.
Only two transitions ((12,11)-(17,10) and (12,11)-(19,10)) were excluded
from this comparison since differences with them showed
systematically large deviations reaching 0.01 \rcm.

In Table~\ref{dunham} a set of Dunham coefficients for $^{23}$Na$^{85}$Rb
is listed which describes the whole set of the measured differences for
both isotopomers with a deviation of 0.0030 \rcm\ and a normalized
standard deviation of 0.84. For calculating the eigenvalues for
$^{23}$Na$^{87}$Rb the usual scaling rules were used.
The distribution of the indexes of the Dunham coefficients
indicates that for description of such a large set of vibrational
and rotational quantum numbers the parameters of the Dunahm expansion
lose their original meaning of ``spectroscopic constants''. The extremely
small values of some coefficients ($\sim 10^{-40}$ \rcm\ !) indicate that they
should be used with caution due to roundoff errors. At last, the extrapolation properties
of such set of Dunham coefficients are doubtful. Therefore, we prefer
to give the full description of the experimental data by a potential
energy curve, which is obviously a much better physical model, and to
present Dunham coefficients only for convenience since there are
still applications where their use is easier and faster.

\begin{table*}
  \centering
  \caption{Dunham coefficients for $^{23}$Na$^{85}$Rb in \rcm,
derived from the experimental data with $v''\le 76$ and $J''\le 200$.}
\vspace{3mm}
\begin{tabular*}{0.5\linewidth}{@{\extracolsep{\fill}}rrrl}\hline
\hspace{1mm} $i$ & $k$\hspace{15mm} & $Y_{ik}$ & \hspace{20mm} \\ \hline
\end{tabular*}
\begin{tabular*}{0.5\linewidth}{@{\extracolsep{\fill}}rrdl}
    1  &  0 &   106.8544964 & \\
    2  &  0 &   -0.37984464 & \\
    3  &  0 &   -7.703864 & $\times 10^{-4}$ \\
    5  &  0 &   -5.390011 & $\times 10^{-7}$ \\
    7  &  0 &   6.5551619 & $\times 10^{-10}$ \\
    9  &  0 &   -1.9725192 & $\times 10^{-12}$ \\
   10  &  0 &   9.9663375 & $\times 10^{-14}$ \\
   11  &  0 &   -2.5123 & $\times 10^{-15}$ \\
   12  &  0 &   3.7418 & $\times 10^{-17}$ \\
   13  &  0 &   -3.33869 & $\times 10^{-19}$ \\
   14  &  0 &   1.652 & $\times 10^{-21}$ \\
   15  &  0 &   -3.48 & $\times 10^{-24}$ \\
    0  &  1 &   7.019021 & $\times 10^{-2}$ \\
    1  &  1 &   -2.943389 &$\times 10^{-4}$ \\
    2  &  1 &   -1.699605 &$\times 10^{-6}$ \\
    4  &  1 &   -1.474183 &$\times 10^{-9}$ \\
    6  &  1 &   5.732233 &$\times 10^{-12}$ \\
    7  &  1 &   -3.68622 &$\times 10^{-13}$ \\
    8  &  1 &   1.031766 &$\times 10^{-14}$ \\
    9  &  1 &   -1.2551 &$\times 10^{-16}$   \\
   11  &  1 &   1.22142 &$\times 10^{-20}$   \\
   13  &  1 &   -1.6097 &$\times 10^{-24}$   \\
   14  &  1 &   9.446 &$\times 10^{-27}  $ \\
    0  &  2 &   -1.20357 &$\times 10^{-7}$   \\
    1  &  2 &   -7.98199 &$\times 10^{-10}$   \\
    2  &  2 &   -4.21837 &$\times 10^{-12}$   \\
    7  &  2 &   -1.2888117 &$\times 10^{-17}$   \\
    8  &  2 &   1.760426295 &$\times 10^{-18}$   \\
    9  &  2 &   -1.17937534 &$\times 10^{-19}$   \\
   10  &  2 &   4.906556 &$\times 10^{-21}  $ \\
   11  &  2 &   -1.3413739 &$\times 10^{-22}$   \\
   12  &  2 &   2.3980927 &$\times 10^{-24} $  \\
   13  &  2 &   -2.684354 &$\times 10^{-26} $  \\
   14  &  2 &   1.70 &$\times 10^{-28}$   \\
   15  &  2 &   -4.6349 &$\times 10^{-31}$   \\
    0  &  3 &   1.4803 &$\times 10^{-13} $  \\
    1  &  3 &   1.9481 &$\times 10^{-15} $  \\
    3  &  3 &   -9.3543 &$\times 10^{-18}$   \\
    8  &  3  &  7.87129 &$\times 10^{-24}$   \\
    9  &  3  &  -6.789833 &$\times 10^{-25}$   \\
   10  &  3  &  2.13759 &$\times 10^{-26} $  \\
   11  &  3  &  -2.61852 &$\times 10^{-28}$   \\
   13  &  3  &  1.974 &$\times 10^{-32}  $ \\
   15  &  3  &  -1.0266 &$\times 10^{-36}$   \\
    1  &  4  &  -1.953 &$\times 10^{-20} $  \\
    2  &  4  &  -1.906 &$\times 10^{-21} $  \\
    5  &  4  &  1.11596 &$\times 10^{-24}$   \\
    6  &  4  &  -7.344 &$\times 10^{-26} $  \\
    8  &  4  &  7.4791 &$\times 10^{-29} $  \\
   10  &  4  &  -6.261 &$\times 10^{-32} $  \\
   11  &  4  &  8.982 &$\times 10^{-34}  $ \\
   14  &  4  &  -6.3065 &$\times 10^{-40}$   \\
    7  &  5  &  -1.3005 &$\times 10^{-33}$   \\
\hline
\end{tabular*}
\label{dunham}
\end{table*}

\section{Results and Discussion}
\label{concl}

In this paper a spectroscopic study of the NaRb ground state is
presented. A variety of laser sources was used to excite several
band systems in both isotopomers of the NaRb molecule and the
induced fluorescence was analyzed with a high resolution
Fourier transform spectrometer. More than 8800 transitions
were identified resulting in a set of more than 4000 ground state
energy levels with wide range of vibrational and rotational
quantum numbers.
An accurate potential energy curve was fitted to the experimental
data reproducing differences between ground state energy levels with
$\sigma=0.0031$ \rcm\ and $\bar{\sigma}=0.70$.

In order to check the consistency of the derived potential
energy curve we performed additional fits in several different ways.
Instead of fitting differences, we applied the commonly used
approach of fitting directly transition frequencies, getting the term
energies of the excited state levels as free parameters. In this case
two representations of the potential were tried: a pointwise and an
analytic form described in detail in Ref.~\cite{Samuelis:00}. Both
representations gave the same quality of the fit compared to the
pointwise potential determined from differences and the potentials were
practically identical in the studied range. Thus no additional parameters
sets are given here in order to avoid the confusion of the reader
what potential form should be preferred.

The classical turning point of the last observed energy level
($v''=76$, $J''=27$) is around 12.4 \AA. Although this point
lies beyond the Le Roy radius for the NaRb 3$^2$S+5$^2$S
asymptote  ($R_{\mathrm {LeRoy}}=11.2$ \AA), the analysis of the PEC
uncertainties indicates that with the present body of experimental 
data it is unsafe to determine $C_6$ and the other dispersion 
coefficients. For this reason the experimental PEC was smoothly 
connected to a long range potential formed by dispersion terms taken
from the literature \cite{Derevianko:01a,Porsev:03}. For better
description of the near asymptotic region of the PEC
transition frequencies to weakly bound energy levels are
needed (see for example \cite{MC:03}). Some improvement might be 
expected already from the
combined analysis of the X$^1\Sigma^+$ and a$^3\Sigma^+$ states
at long internuclear distances which is in progress.

Weiss et al. \cite{Weiss:03} applied the MLJ potential from \cite{Docenko:02}
and extended it for $R>11$ \AA\  by the long range parameters from
\cite{Marinescu:99,Derevianko:01a} and the exchange interaction taken from
\cite{Smirnov:65} to determine the singlet and triplet scattering lengths
for 3$^2$S Na + 5$^2$S Rb. Using the newly derived potential for the inner
part we constructed the full potential with the same long range parameters
as in \cite{Weiss:03} and derived a singlet scattering
length of +305a$_0$ ($a_0\approx 0.52918$ \AA, the Bohr radius)
for $^{23}$Na$^{85}$Rb and  +103a$_0$ for $^{23}$Na$^{87}$Rb.
These results do not agree with the ones from Ref. \cite{Weiss:03}: +167a$_0$ (+50a$_0$/-30a$_0$) for $^{23}$Na$^{85}$Rb
and 55a$_0$ (+3a$_0$/-3a$_0$) for $^{23}$Na$^{87}$Rb. We used exactly the same
approach for the long range as in \cite{Weiss:03} to see clearly the influence
of the inner part of the potential. By the new values we will not claim
that we have determined more realistic scattering lengths. Rather 
we want to point out that the
existing data do not allow to specify the scattering length in an
useful uncertainty interval. Note also that slight changes in the exchange energy
still allow to alter the sign of the scattering length
in $^{23}$Na$^{85}$Rb. Thus energy levels closer to the asymptote
($< 4.5$ \rcm) must be studied.

\section{Acknowledgments}

This work is supported by DFG through SFB 407.
The authors appreciate the assistance of M. Klug and Ch. Samuelis
during the experiments. A.P. gratefully acknowledges the research
stipend from the Alexander von Humboldt Foundation. O. D., M. T. and
R. F. acknowledge the support by the University of Latvia, by 
the Latvian Science Council grant 01.0264, the Latvian Ministry 
of Education and Science (grant TOP 02-45)
as well as by NATO SfP 978029 - Optical Field Mapping grant.


\begin{thebibliography}{10}

\bibitem{Shaffer:99}
James~P. Shaffer, Witek Chalupczak, and N.~P. Bigelow.
\newblock {\em Phys. Rev. A} 60, R3365 (1999).

\bibitem{Ferrari:02}
{G. Ferrari, M. Inguscio, W. Jastrzebski, G. Modugno, G. Roati and
  A. Simoni}.
\newblock {\em Phys. Rev. Lett.} 89, 53202  (2002).

\bibitem{Hadzibabic:02}
{Z. Hadzibabic, C. A. Stan, K. Dieckmann, S. Gupta, M. W. Zwierlein,
A. G\"orlitz and W. Ketterle}.
\newblock {\em Phys. Rev. Lett.} 88, 160401 (2002).

\bibitem{Weiss:03}
{S. B. Weiss, M. Bhattacharya and N. P. Bigelow}.
\newblock {\em {arXiv:physics/0303006 v1}}.

\bibitem{Wang2:91}
Y.-C. Wang, M.~Kajitani, S.~Kasahara, M.~Baba~K. Isikawa, and H.~{Kat\^{o}}.
\newblock {\em J. Chem. Phys.} 95, 6229 (1991).

\bibitem{Kasahara:96}
S.~Kasahara, T.~Ebi, M.~Tanimura, H.~Ikoma~K. Matsubara, M.~Baba, and
  H.~{Kat\^{o}}.
\newblock {\em J. Chem. Phys.} 105, 1341 (1996).

\bibitem{Docenko:02}
{O. Docenko, O. Nikolayeva, M. Tamanis, R. Ferber, E. A. Pazyuk and
  A. V. Stolyarov}.
\newblock {\em Phys. Rev. A} 66, 052508 (2002).

\bibitem{Marinescu:99}
M.~Marinescu and H.~R. Sadeghpour.
\newblock {\em Phys. Rev. A} 59, 390 (1999).

\bibitem{Zemke:NaRb}
W.~T. Zemke and W.~C. Stwalley.
\newblock {\em J. Chem. Phys.} 114, 10811 (2001).

\bibitem{Korek:00}
{M. Korek, A.R. Allouche, M. Kobeissi, A. Chaalan, M. Dagher,
K. Fakherddin and M. Aubert-Fr\'econ}.
\newblock {\em Chem. Phys.} 256, 1 (2000).

\bibitem{Zaitsevskii:01}
{A. Zaitsevskii, S.O. Adamson, E.A. Pazyuk, A.V. Stolyarov, O.
  Nikolayeva, O. Docenko, I. Klincare, M. Auzinsh, M. Tamanis,
  R. Ferber and R. Cimiraglia}.
\newblock {\em Phys. Rev. A} 63, 052504 (2001).

\bibitem{Allard:02}
{O. Allard, A. Pashov, H. Kn\"ockel and E. Tiemann}.
\newblock {\em Phys. Rev. A} 66, 42503 (2002).

\bibitem{Wang:92}
Y.-C. Wang, K.~Matsubara, and H.~{Kat\^{o}}.
\newblock {\em J. Chem. Phys.} 97, 811 (1992).

\bibitem{Takahashi:81}
N.~Takahashi and H.~{Kat\^{o}}.
\newblock {\em J. Chem. Phys.} 75, 4350 (1981).

\bibitem{Nikolayeva:00}
{O. Nikolayeva, I. Klincare, M. Auzinsh, M. Tamanis, R. Ferber,
  E. A. Pazyuk, A. V. Stolyarov, A. Zajtsevskii and R. Cimiraglia}.
\newblock {\em J. Chem. Phys.} 113, 4896 (2000).

\bibitem{ipaasen}
A.~Pashov, W.~{Jastrz\c{e}bski}, and P.~Kowalczyk.
\newblock {\em Comput. Phys. Commun.} 128, 622 (2000).

\bibitem{Numer}
W.~H. Press, S.~A. Teukolski, W.~T. Vetterling, and B.~P. Flannery.
\newblock {\em {Numerical Recipes in Fortran 77}}.
\newblock Cambridge Unversity Press, 1992.

\bibitem{Derevianko:01a}
{A. Dervianko, J. F. Babb and A. Dalgarno}.
\newblock {\em Phys. Rev. A} 63, 052704 (2001).

\bibitem{Porsev:03}
{S. G. Porsev and A. Derevianko}.
\newblock {\em J Chem. Phys.} 119, 844 (2003).

\bibitem{MC:03}
{O. Allard, C. Samuelis, A. Pashov, H. Kn\"ockel and E. Tiemann}.
\newblock {\em Eur. Phys. J. D} 26, 155 (2003).

\bibitem{Li2F:00}
A.~Pashov, W.~{Jastrz\c{e}bski}, and P.~Kowalczyk.
\newblock {\em J. Chem. Phys.} 113, 6624 (2000).

\bibitem{Wilkinson:SVD}
J.~H. Wilkinson and C.~Reinsch.
\newblock {\em Linear Algebra}.
\newblock Springer-Verlag, Berlin, 1971.

\bibitem{Samuelis:00}
C.~Samuelis, E.~Tiesinga, T.~Laue, M.~Elbs, H.~{Kn\"{o}ckel}, and E.~Tiemann.
\newblock {\em Phys. Rev. A} 63, 012710 (2000).

\bibitem{Smirnov:65}
{B. M. Smirnov and M. I. Chibisov}.
\newblock {\em {JETP}} 21, 624 (1965).

\end{thebibliography}

 \end{document}